\documentclass[
aip,
rsi,
amsmath,
amssymb,
reprint
]{revtex4-1}

\usepackage{graphicx}
\graphicspath{{fig/}}
\usepackage{epstopdf}

\usepackage{dcolumn}
\usepackage{bm}
\usepackage{natbib}

\draft 

\begin{document}

\preprint{AIP/123-QED}

\title{An Atom Trap Trace Analysis System for Measuring Krypton Contamination in Xenon Dark Matter Detectors }

\author{E. Aprile}
\author{T. Yoon}
\email[]{ty2182@columbia.edu}
\author{A. Loose}
\author{L. W. Goetzke}
\author{T. Zelevinsky}

\affiliation{Department of Physics, Columbia University, New York, NY 10027-5255, USA}

\date{\today}

\begin{abstract}
We have developed an atom trap trace analysis (ATTA) system to measure Kr in Xe at the part per trillion (ppt) level, a prerequisite for the sensitivity achievable with liquid xenon dark matter detectors beyond the current generation. Since Ar and Kr have similar laser cooling wavelengths, the apparatus has been tested with Ar to avoid contamination prior to measuring Xe samples. A radio-frequency (RF) plasma discharge generates a beam of metastable atoms which is optically collimated, slowed, and trapped using standard magneto-optical techniques. Based on the measured overall system efficiency of $1.2 \times 10^{-8}$ (detection mode) we expect the ATTA system to reach the design goal sensitivity to ppt concentrations of Kr in Xe in $<2$ hours.
\end{abstract}


\maketitle

\section{Introduction}

Large volume liquid xenon (LXe) detectors are leading the field of dark matter direct detection with the best sensitivity achieved by the XENON100 experiment.\cite{PhysRevLett.109.181301} A two order of magnitude improvement in sensitivity is projected by the next generation experiment, XENON1T~\cite{Aprile:2012zx}, which will use more than three tons of Xe as target. Xe is extracted from the atmosphere with a typical krypton (Kr) contamination at the part per million (ppm) level. The Kr contamination contributes background events through the radioactive isotope $^{85}$Kr, which undergoes $\beta$-decay with a 687 keV end point and 10.8 year half-life. This background from $^{85}$Kr strongly limits the detection sensitivity of LXe dark matter detectors. The sensitivity reach of XENON1T requires a  Kr/Xe contamination below a part per trillion (ppt). Cryogenic distillation is an established technology for Xe purification from Kr at the ppt level. However, practical and fast measurements at or below such extremely low levels of contamination are beyond the scope of conventional methods such as low level decay counting~\cite{doi:10.1146/annurev.ns.25.120175.002231}, mass spectrometers with liquid nitrogen cold traps~\cite{Dobi20111}, or accelerator mass spectroscopy.\cite{doi:10.1146/annurev.ns.30.120180.002253} Hence, we have proposed the development of the atom trap trace analysis (ATTA) system described in this paper.

ATTA systems are based on laser cooling, trapping, and counting of single atoms.\cite{Chen05111999,McKinsey2005524,RevModPhys.84.175} The extraordinary selectivity of the ATTA method is a result of the high number of resonant photon-atom interactions that are used to slow and capture an atom. Detection of other atomic or molecular species can be excluded at the 90\% confidence level even at parts per quadrillion sensitivity.\cite{PhysRevLett.106.103001} The photons are generated by a narrow bandwidth laser which is tuned to an isotope-specific optical transition. The setup which we have developed will allow for a rapid and reliable measurement of Kr concentrations in Xe at the ppt level, by trapping the most abundant (57\%) isotope $^{84}$Kr. We can then infer the $^{85}$Kr concentration since the isotopic abundance of $^{85}$Kr ($1.5\times 10^{-11}$) is known.\cite{Chen05111999} A desired contamination of less than one $^{85}$Kr in $10^{23}$ Xe atoms thus corresponds to a ppt level contamination of $^{84}$Kr in Xe. Ultra-sensitive Kr trace analysis is also used for geological dating and studies of transport processes in the atmosphere, oceans, and groundwater~\cite{annurevnucl2004,GeophysResLett2003,GeophysResLett2003,PhysRevLett.106.103001}, monitoring nuclear-fuel reprocessing activities, and may be of interest for the rare gas industry to guarantee the purity of gases used in a variety of applications.

This paper is organized as follows. In Sec. \ref{sec:Apparatus}, we give a description of our ATTA device, which consists of the laser and vacuum subsystems, the radio-frequency (RF) discharge source, and a single atom detection setup. In Sec. \ref{sec:Characterization}, we experimentally characterize the parameters which determine the overall system efficiency, and demonstrate the ability to detect single atoms with a good signal to noise ratio. Finally, in Sec. \ref{sec:Discussion} we describe the system performance for a Xe carrier gas containing ppm Kr and show that ppt sensitivity of $^{84}$Kr detection is expected in $\sim 1.5$ hours of measurement time, with further improvements underway.

\section{Experimental Apparatus}
\label{sec:Apparatus}

\begin{figure}
\begin{center}
\includegraphics*[width=.48\textwidth]{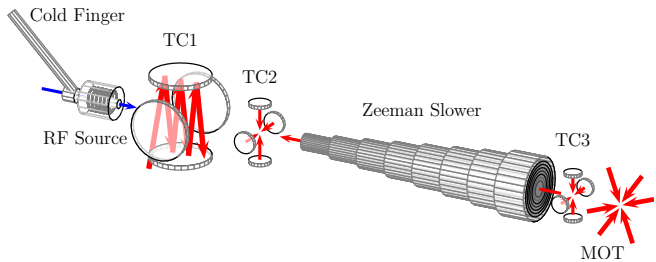}
\caption{ATTA system schematic. Selected components and laser beam directions (red arrows) are shown. Metastable atoms are generated and cooled by a RF source with an attached cold finger, collimated by three transverse cooling stages (TC1-TC3), decelerated with the Zeeman slowing technique, and captured in the magneto-optical trap (MOT) for detection.}
\label{fig:atta_sketch}
\end{center}
\end{figure}

A simplified sketch of the ATTA system is shown in Fig.~\ref{fig:atta_sketch}. The light for the atomic beam slowing, transverse cooling, and the magneto-optical trap (MOT) is generated by a semiconductor-based laser system. The gas sample is expanded into an ultra-high vacuum system with a base pressure $<10^{-9}$ torr for cooling and trapping. During gas flow, the pressure  in the source chamber is maintained at 0.4 mtorr, while the pressure in the MOT chamber remains at $10^{-8}$ torr. A beam of metastable atoms is created by passing the injected gas sample through a RF discharge region, where inelastic collisions with electrons and ions form a plasma and excite the ground state atoms. The atomic beam is collimated by two-dimensional optical molasses, slowed by radiation pressure with Zeeman tuning, and captured in the MOT.  The metastable atoms that are periodically trapped in the MOT are detected and counted by their flourescence with an avalanche photodiode (APD). For a known Xe gas inflow rate and overall system efficiency, the number of detected metastable $^{84}$Kr* atoms per unit time can be used to determine the $^{85}$Kr/Xe ratio.

To avoid system contamination with Kr, the apparatus has been set up and tested with $^{40}$Ar* before transitioning to $^{84}$Kr* in Xe. There are two main sources of contamination at ultra-low Kr concentrations. The first is outgassing from the walls of the vacuum system. This is however not a significant source of background for ppt-level measurements, at typical gas consumption rates on the order of $10^{16}$ atoms/s.~\cite{McKinsey2005524} No gas recirculation is used and only metastable atoms are detected, therefore the residual Kr downstream from the source chamber does not affect the measurements. The second source of sample contamination is Kr implanted in the walls of the RF discharge source during operation, and becomes important at very low Kr concentrations.~\cite{NatureSciRep2013} We did not observe this effect at ppm-level measurements of Kr in Xe, where we operated the discharge with Ar for re-alignment between different Xe samples.  However, we expect that the system must be operated with pure Ar for an extended period of time before switching from reference samples with ppm or part per billion (ppb) Kr abundance to the ppt-level samples of interest.~\cite{GeophysResLett2003,NatureSciRep2013}

\subsection{Laser System}

\begin{figure}
\begin{center}
\includegraphics*[width=.48\textwidth]{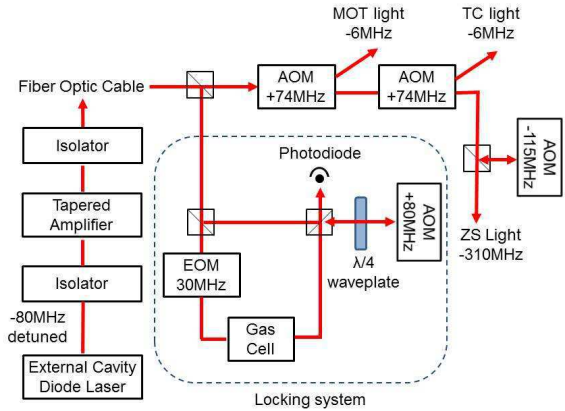}
\caption{Solid-state laser system schematic. An external cavity diode laser (ECDL) is locked to a gas reference cell using saturated absorption spectroscopy. After amplification the laser light is distributed to the transverse cooling (TC) stages, magneto-optical trap (MOT), and Zeeman slower (ZS) with appropriate intensities, polarizations, and frequency detunings. AOM: acousto-optical modulator; EOM: electro-optical modulator.}
\label{fig:laser_system}
\end{center}
\end{figure}

Both $^{40}$Ar and $^{84}$Kr have no hyperfine structure, and can be cooled and trapped without the use of optical repumping. A semiconductor laser setup supplies the laser beams used for the transverse cooling stages, Zeeman slowing, and the MOT, as shown in Fig.~\ref{fig:laser_system}. The external cavity diode laser (ECDL) with an output power of 50 mW is custom-built in the Littman configuration (diode: Sacher SAL-0840-060). The laser output is amplified by a tapered amplifier (Eagleyard EYP-TPL-0808-01000) and delivered to the optical table via a high-power polarization maintaining optical fiber.

Approximately 20 mW of the laser power after the fiber is split off for the locking system.  The Doppler-free heterodyne spectroscopy technique is used to lock the ECDL to the required wavelength, with an offset of $-80$ MHz. For this purpose, a strong pump beam is detuned by $+160$ MHz and amplitude modulated by double-passing an acousto-optical modulator (AOM), and a weak probe beam is phase modulated by a resonsant electro-optical modulator (EOM). A lock-in amplifier processes the signal detected by a photodiode. This provides the error signal for a piezo driver controlling the ECDL grating. The emission wavelengths needed for the closed cooling transitions are 811.7542 nm for $^{40}$Ar* (4$^3$P$_2$ -- 4$^3$D$_3$) and 811.5132 nm for $^{84}$Kr* (5$^3$P$_2$ -- 5$^3$D$_3$). The natural linewidths of both atomic transitions are $\sim 2\pi \times 6$ MHz.

The main portion of the laser power after the fiber is used for slowing, cooling and trapping. An AOM shifts the frequency of the laser by \mbox{$+74$ MHz}, and this first-order beam is used for magneto-optical trapping. The unchanged zero-order beam is shifted by $+74$ MHz with the second AOM and is used for the transverse cooling stages. The unshifted beam double-passes a $-115$ MHz detuned AOM for an effective detuning of $-310$ MHz. This beam propagates along the Zeeman slower axis towards the source, and slows the atomic beam.

The optical coupling into the tapered amplifier and fiber optics introduces laser intensity fluctuations, which limit the signal to noise ratio of single atom detection. To minimize these fluctuations, a custom-made proportional-integral-derivative (PID) control circuit with a photodiode stabilizes the laser intensity of the trapping beams by feedback to the MOT AOM.

\subsection{Vacuum System}

\begin{figure}
\begin{center}
\includegraphics*[width=.48\textwidth]{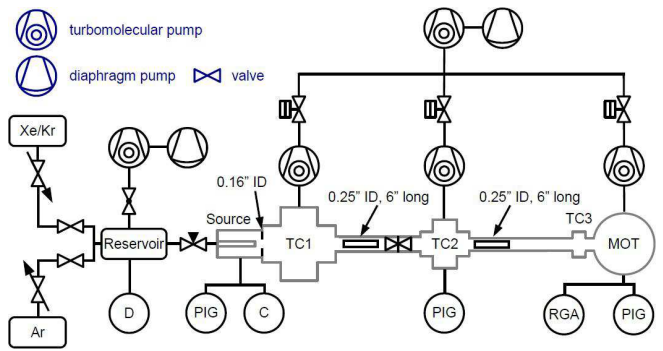}
\caption{Vacuum system schematic. The reservoir chamber is connected to the source chamber via an ultra-fine leak valve. Differential pumping stages ensure a sufficient pressure gradient between source chamber, transverse cooling chambers (TC1-TC3), and detection chamber (MOT). The vacuum is characterized by a pressure transducer (D), capacitive manometer (C), combined Pirani/cold cathode gauges (PIG), and a residual gas analyzer (RGA).}
\label{fig:vacuum_system}
\end{center}
\end{figure}

The vacuum system consists of the following main components: sample reservoir, source chamber, three transverse cooling chambers, and detection chamber. A simplified sketch is shown in Fig.~\ref{fig:vacuum_system}. Three turbopumps, backed by a single turbopump station with a dry diaphragm pump, differentially pump the system. A separate turbopump station evacuates the reservoir chamber. A manual all-metal ultra-fine leak valve separates the reservoir chamber from the source chamber, and allows for the pressure in the source chamber to be maintained at about 0.4 mtorr during operation. A gate valve separates the first and second transverse cooling chambers.  Pneumatic valves on the outputs of the three main turbopumps allow for breaking vacuum to the backing pumping station while maintaining vacuum in the system, and prevent the loss of vacuum in case of a power failure. Pirani and cold-cathode gauges monitor the base pressure in the reservoir chamber, source chamber, second transverse cooling chamber, and detection chamber. The base pressure is maintained at $<10^{-8}$ torr in all chambers.

During measurement, part of the gas sample is transferred to the reservoir chamber, which has a volume of 0.63 l. Typical sample volumes are $\sim$1 l STP. A pressure transducer (with a quoted accuracy of 0.1\% of full scale) attached to the reservoir is used to measure the total sample size and gas consumption rate. A capacitive manometer is used to monitor the pressure in the source chamber with a quoted accuracy of 0.5\%, as the discharge efficiency is relatively sensitive to the source pressure. The inner surface of the detection chamber is painted black using ultra-high vacuum compatible paint (AZ Technology, MLS-85SB) to minimize scattering of the cooling and trapping beams. A residual gas analyzer connected to the detection chamber is used to monitor the vacuum composition.

\subsection{RF Discharge Source}

A beam of atoms is generated by expanding the gas sample into the vacuum system. Efficient ultraviolet laser sources to drive noble gas cooling transitions from the ground state are currently not available, hence metastable atoms with optically accessible transitions have to be generated. A RF discharge source ignites and sustains a low-pressure plasma in the presence of the gas flow. Inelastic collisions with the plasma electrons and ions excite a fraction of the effusing atoms into the long-lived metastable state.\cite{chen:271}

\begin{figure}
\begin{center}
\includegraphics*[width=.48\textwidth]{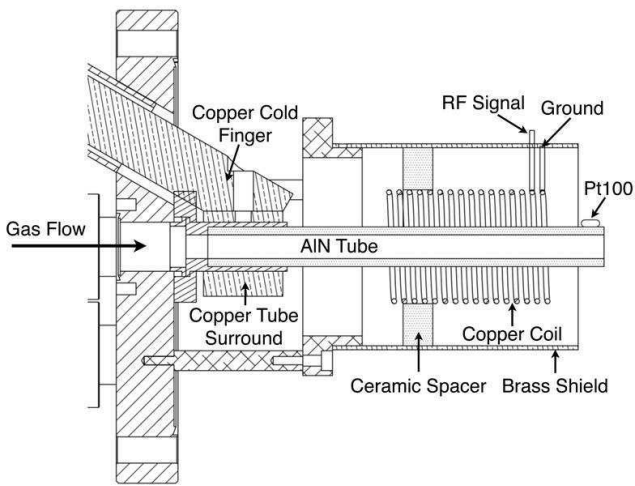}
\caption{Cross-section of the RF source. The gas expands through an AlN tube which is cooled by a cold finger. A plasma is generated by an RF signal applied to the copper coil, producing metastable atoms. Temperature is monitored by the Pt100 platinum resistive temperature sensor.}
\label{fig:source_section_schematic}
\end{center}
\end{figure}

A schematic of the source is shown in Fig.~\ref{fig:source_section_schematic}. The design parameters are chosen based on Refs.~\cite{macalpine1959} and \cite{chen:271}.  The gas flows through an electrically insulating and thermally conductive tube made of aluminium nitride (AlN). To avoid quenching of the metastables via collisions between Xe and $^{84}$Kr*, a differential pumping gasket immediately after the AlN tube exit reduces the gas pressure. A Pt100 temperature sensor is mounted directly on the AlN tube. The brass shield is 11.1 cm long with an inner diameter of 6.35 cm. The copper coil is 7.6 cm long with 16 windings, made from 18 American wire gauge (AWG) bare copper wire, and has an inner diameter of 3.4 cm. Holes were drilled in the brass shield through which the ground and RF power wire are connected. The flexible coil is soldered to the shield at one end, and a Macor ceramic spacer keeps it centered. A copper cold finger connected to a pulse tube refrigerator (Iwatani PDC08) is used to cool the gas to $\sim 160$ K. This increases the fraction of Kr atoms below the capture velocity of the Zeeman slower by more than a factor of three. Cryogenic high-vacuum compatible grease on the connections between the copper pieces increases the cooling efficiency.

The RF signal is generated by a voltage controlled oscillator. It passes through a voltage controlled variable attenuator, and is subsequently amplified by an RF amplifier (Mini-Circuits ZHL-50W-52-S). The amplified signal is fed into the source by a SubMiniature version A (SMA) cable feedline. Care has to be taken to match the impedances of the amplifier and the source to maximize the fraction of the RF power deposited into the plasma, as well as to minimize heating.\cite{sukenik:493}

To ignite the plasma source, the RF power is first increased until a bright discharge is visible through the rear viewport, then the power is reduced. We observe hysteresis between the low-brightness mode at ignition and at lower RF powers, and the high brightness mode at higher RF powers.\cite{rf-discharge-coupling} The metastable atom flux is significantly increased when the source operates in the high-brightness mode; running the discharge at the minimal power required for this mode reduces heating. Typical operating parameters for Ar are 15 W of applied RF power at 118 MHz, at a pressure of 0.6 mtorr and temperature of $160$ K.

\begin{figure}
\begin{center}
\includegraphics*[width=.47\textwidth]{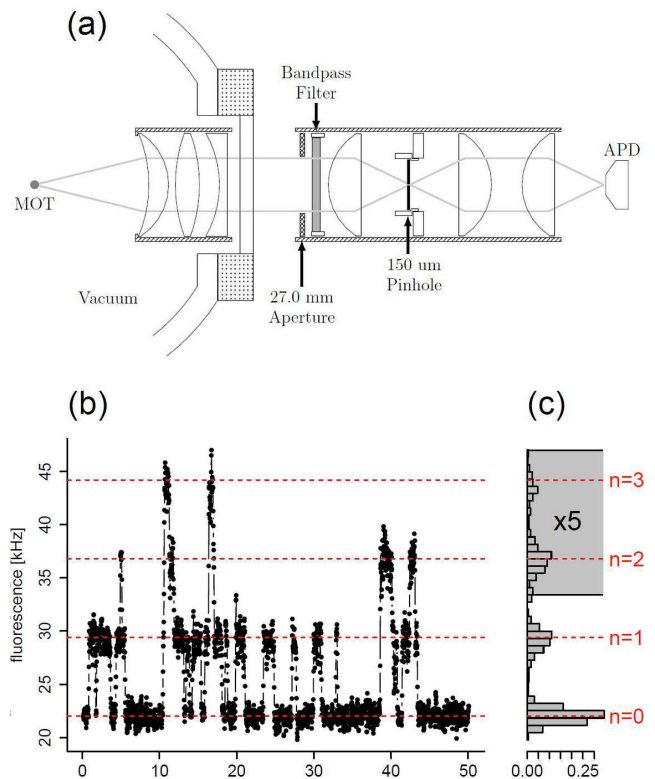}
\caption{(a) Single atom detection, optical setup. A lens tube system images the MOT fluorescence onto the active area of an APD. A customized lens system is mounted in vacuum; a lens tube and APD detector are mounted in a light-tight enclosure. The 150 $\mu$m pinhole acts as a spatial filter, and a band-pass filter rejects background light. (b) Fluorescence signal measured by the APD and (c) its histogram. Discrete steps of $7.4(1)$ kHz are visible. Horizontal lines are guides to the eye. The background from scattered trapping light and atomic beam fluorescence is 22 kHz.}
\label{fig:sad_sketch}
\end{center}
\end{figure}

\subsection{Single Atom Detection Setup}

The low contamination level of Xe by Kr after cryogenic distillation leads to an expected average MOT population of less than one atom. Thus, the system must have the ability to efficiently detect single trapped atoms.

A sketch of the single atom detection setup is shown in Fig.~\ref{fig:sad_sketch}(a). We employ six anti-reflection coated lenses between the MOT and the detector to gather and focus the fluorescence photons. The first three lenses, used for light collection and collimation, are mounted in vacuum. The other three lenses are mounted outside the vacuum in a lens tube. A 150 $\mathrm{\mu}$m pinhole is used for spatial filtering. It has been chosen to be slightly smaller than the typical trapped atom cloud diameter in order to sufficiently reduce background light due to stray reflections from optical components and the chamber. A $810\pm 5$ nm band-pass filter (BP) is used to remove ambient light entering the lens tube. The fluorescence photons are detected by an avalanche photo diode (APD, Perkin-Elmer single photon counting module SPCM-AQRH-12) mounted on a translation stage. Both the lens tube and APD are mounted in a light-tight enclosure. We employ a microcontroller fail-safe circuit to protect the APD from overexposure, using the Arduino open-source electronic prototyping platform.\cite{gordon:1150,Arduino} The fail-safe also relays the transistor-transistor logic (TTL) photon count signals of the APD to a DAQ card (National Instruments PCI Express 6321) connected to a computer. Software records the count frequency.

A single atom trapped in the MOT scatters approximately $10^7$ photons/s into the full solid angle of 4$\pi$ sr. The effective solid angle of our setup for light collection is 1.8\%, as determined by ray-tracing simulations taking into account the (180 $\mu$m)$^2$ detector area. Including reflection losses, detector efficiency and transmittance of the BP (0.58), the ideal fluorescence signal without spatial filtering is 31 kHz per atom.

\section{Characterization Of The Instrument}
\label{sec:Characterization}

\subsection{System Efficiency for Argon}

To estimate the consumption rate for Ar, the reservoir is filled with gas which is then allowed to flow out through the ultra-fine leak valve. The pressure decrease in the reservoir is measured while maintaining the source chamber pressure at 0.6 mtorr. Using the ideal gas law, we estimate the consumption rate to be $6 \times 10^{16}$ atoms/s.

To characterize the production of metastable atoms by the RF source, a fluorescence laser was set perpendicular to the atomic beam direction in the first transverse cooling region. The detuning of the laser beam was varied and the fluorescence light from atoms of different transverse velocity classes was measured with a CCD camera.  The most probable longitudinal velocity of the Ar atoms could be estimated from the transverse velocity and geometrical considerations to be 283 m/s, corresponding to a temperature of 129 K. The design capture velocity of our Zeeman slower is 245 m/s for Kr and 250 m/s for Ar. Assuming the velocity distribution of a one-dimensional thermal beam for the atoms leaving the discharge source, source cooling increases the capture fraction from 18\% for Kr and 6\% for Ar (at 400 K), to 68\% and 33\%, respectively.  A metastable atom flux of $6 \times 10^{11}$ atoms/s with an angular flux density of $9 \times 10^{13}$ s$^{-1}$ sr$^{-1}$ is generated. Ar atoms are excited by the RF discharge source with an efficiency of $\sim 10^{-5}$.

The first transverse cooling stage increases the atom number captured by the MOT by a factor of 20, the second transverse cooling by a factor of 2. The rapid transverse expansion of the slowed atom beam in the 4.5 cm after the Zeeman slower exit reduces the efficiency of the third transverse cooling. With only the horizontal beam of the third transverse cooling stage used, a 20\% increase in the trapped atom number can be observed. Optimizing all experimental parameters while preserving a small trapping beam detuning and high MOT field gradient, allows for $10^7$ atoms to be trapped at a time. Switching off the Zeeman slower field reduces the atom number in the MOT to $<10^3$.

To determine the loading rate of the MOT, a CCD camera is set up at one of the viewports of the MOT chamber to detect the fluorescence of the atom cloud. On another viewport, a photodiode is mounted to provide increased temporal resolution. The brightness of the CCD image and the photocurrent generated by the photodiode are converted to the number of photons emitted from the trapped atoms. From this, the MOT size and the number of trapped atoms, $N(t)$, are derived. The trap parameters are modeled by
\begin{equation}
\frac{dN(t)}{dt}=L-\frac{N(t)}{\tau}- \beta n N(t)
\end{equation}
with loading rate $L$, quadratic collision term $\beta$, MOT lifetime $\tau$, and constant density $n$.\cite{Steane1992} Fitting Eq. (1) for trap loading and decay measurements, we determined a loading rate of $1.8 \times 10^8$ atoms/s. Given our consumption rate, the overall efficiency of the setup is $3 \times 10^{-9}$ for $^{40}$Ar* in Ar.

\subsection{Single Atom Detection}

To demonstrate fluorescence detection with single atom resolution, the efficiency of the setup was reduced by blocking the Zeeman slower laser beam as well as the transverse cooling light. Thus, only a few atoms at a time are captured in the MOT. Figures ~\ref{fig:sad_sketch} (b,c) show a plot of the APD count rate over 50 seconds with an integration time of 60 ms, as well as the associated histogram. Discrete steps of $7.4(1)$ kHz are visible in the photon count rate. These discrete signal levels correspond to zero, one, two, and three trapped atoms in the MOT, and the single atom signal to noise ratio is 5. The atomic beam fluorescence and stray reflections of the trapping laser beams contribute a background of 22 kHz, while the dark count of the APD module is negligible at 0.3 kHz. The additional Zeeman slower laser beam contribution to the APD background during the Kr measurements is compensated by the absence of the atomic beam fluorescence. The average lifetime of a single metastable Ar atom trapped in the MOT is $0.7(1)$ seconds.

The MOT magnetic field as well as the detuning of the trapping beams were not changed for detection, while the trapping beam intensity was reduced by a neutral density filter with a transmission of 0.2. Using the full trapping beam intensity, while increasing the loading rate by a factor of $\sim 20$, also increases the APD background significantly. We plan to implement switching of the beam intensity between a loading and detection mode, as we expect an order of magnitude increase in system efficiency. For the following measurements we used the reduced trapping beam intensity.

\subsection{System Efficiency for Krypton in Xenon}

We found that both the optimal RF power and the operating pressure of the Xe discharge are lower than for Ar. We typically operate the RF source at a pressure of 0.4 mtorr with 7 W of applied RF power. The consumption rate with the maximum metastable atom flux is $2.5 \times 10^{16}$ atoms/s.

To characterize the system efficiency for Kr in Xe carrier gas, we admixed well-defined amounts of Kr to 10-ppb level pure Xe. At 12 and 0.8 ppm Kr contaminations, the system efficiency had to be reduced in order to capture less than one atom at a time, even when using the reduced trapping beam intensity. For this purpose, the transverse cooling stages were disabled, and the Zeeman slower beam diameter was reduced. The single atom counts measured by the APD indicate an expected $^{84}$Kr* loading rate of $\sim 1.8\times10^{-4}$/s for ppt-level Kr in Xe, using the detection-mode trapping beam. This corresponds to an average of one trapped atom every 90 minutes. We expect to test the system with a sample of Xe gas from the XENON100 experiment as soon as the ongoing dark matter search is terminated. The sample will have Kr/Xe at the ppt level as measured by a rare gas spectrometer.~\cite{lindemann:13084806}

Comparing the known loading rate for Ar in trapping mode to the measured Kr loading rate in detection mode, with the $^{84}$Kr abundance of 0.57, we estimate an increase of trapping efficiency $\rho = 35$ for Kr in Xe in detection mode, using the relation $1.8 \times 10^{8}\mathrm{ /s} \times 1/20 \times 10^{-12} \times 0.57 \times \rho = 1.8 \times 10^{-4}\mathrm{/s}$. With the lower consumption rate of Kr in Xe taken into account, the overall system efficiency increases by a factor of 84. The theoretical efficiency increase of the Zeeman slower due to the lower thermal velocity of Kr and the Kr-optimized design accounts for a factor of two. We expect that the remainder is mostly due to an increased efficiency in the production of Kr* in Xe by the RF source. To the best of our knowledge, the role of Xe as carrier gas in the production of metastable Kr atoms in a RF discharge has not yet been investigated. However, a systematic study using Ne, Kr and Xe as carrier gases for Ar has been conducted,~\cite{rudinger:036105} where Xe yielded the optimal fractional metastable atom population of $2\times 10^{-4}$.

Table~\ref{table:loading_rate} summarizes the performance data of the complete system when tested with $^{40}$Ar* and with Xe containing ppm-level $^{84}$Kr*, as well as the loading rate enhancement factors due to the transverse cooling stages.

\begin{table}
    \begin{tabular}{r l}
        \hline
       gas consumption rate: &  $6 \times 10^{16}$ atoms/s (Ar) \\
        &  $2.5 \times 10^{16}$ atoms/s (Xe) \\
       metastable atom flux: &  $6 \times 10^{11}$ atoms/s (Ar)\\
       angular flux density: &  $9 \times 10^{13}$ atoms/s/sr (Ar) \\
       source efficiency:  &  $10^{-5}$ (Ar*/Ar)  \\
       &  $10^{-4}$ (Kr*/Kr in Xe)  \\
       MOT loading rate: & $1.8 \times 10^{8}$ atoms/s (Ar)\\
       overall efficiency:& $3 \times 10^{-9}$ (Ar, trapping mode)\\
                                & $1.2 \times 10^{-8}$ (Kr, detection mode)\\
       \hline
       flux enhancement: &TC1: 20$\times$; TC2: 2$\times$;\\
        & TC3 (horizontal): 1.2$\times$ (Ar)\\
        \hline
    \end{tabular}
\caption{Performance data of the ATTA device for $^{40}$Ar* in Ar and for Xe containing ppm-level $^{84}$Kr*.}
\label{table:loading_rate}
\end{table}

\section{Discussion}
\label{sec:Discussion}

We have constructed, characterized, and optimized the Kr in Xe ATTA system using Ar, to avoid contamination by Kr.  For $^{84}$Kr* in a Xe carrier, the overall system efficiency is almost two orders of magnitude higher than for $^{40}$Ar* in Ar.  In detection mode, we expect to trap 0.7 $^{84}$Kr* atoms per hour at a ppt-level contamination.  With switching the trapping beam intensity between trapping and detection modes, we expect an additional increase in overall efficiency exceeding an order of magnitude.  Raising the efficiency of the transverse cooling stages by increasing the available laser power and installing a two-dimensional MOT in the second transverse cooling zone should further improve the counting rate, and therefore the system sensitivity.

\begin{acknowledgments}
We thank the National Science Foundation and Columbia University for the support of this experiment with the Major Research Instrumentation award No. 0923274.

We are grateful to Z.-T. Lu and colleagues at Argonne National Laboratory for advice and useful discussions. We thank C. Allred for contributions to the initial stages of the project, and G. Reinaudi for valuable assistance.
\end{acknowledgments}
%
\end{document}